\documentclass{IEEEtran}
\usepackage{url}
\usepackage[utf8]{inputenc}
\usepackage{color} 
\usepackage{t1enc}  
\usepackage{graphicx}
\usepackage{color}   
\usepackage{amsmath} 
\usepackage{amsfonts}   
\usepackage{enumerate}  
\usepackage{longtable}
\usepackage{dcolumn}
\usepackage{bm}
\usepackage{epsfig}
\usepackage{hyperref}
\usepackage{subfigure}
\usepackage{wrapfig}
\usepackage{bbm}

\newcommand{\Eff}{\text{Eff}}
\newcommand{\CDF}{\text{CDF}}

\title{Temporal influence over the Last.fm social network%
\thanks{Research supported in part by the EC FET Open project
``New tools and algorithms for directed network analysis''
(NADINE No 288956) and by the grant OTKA NK 105645.
The work of Robert Palovics reported in this paper has been developed in the framework of the
project ``Talent care and cultivation in the scientific workshops of BME''
project. This project is supported by the grant TAMOP -
4.2.2.B-10/1--2010-0009.
Work conducted at the E\"otv\"os University, Budapest was partially supported by the
European Union and the European Social Fund through project FuturICT.hu (grant no.: 
TAMOP-4.2.2.C-11/1/KONV-2012-0013). The research was carried out as part of the EITKIC\_12-1-2012-0001 project,
which is supported by the Hungarian Government, managed by the National
Development Agency, financed by the Research and Technology Innovation Fund
and was performed in cooperation with the EIT ICT Labs Budapest Associate
Partner Group. (\url{www.ictlabs.elte.hu})}}

\author{\IEEEauthorblockN{R\'obert P\'alovics$^{1,2}$ \quad Andr\'{a}s A.\ Bencz\'{u}r$^{1,3}$}\\
\IEEEauthorblockA{$^1$Institute for Computer Science and Control, Hungarian Academy of Sciences (MTA SZTAKI)\\
$^2$Technical University Budapest \\
$^3$E\"otv\"os University Budapest \\
\texttt{\{rpalovics,\,benczur\}@ilab.sztaki.hu}}
}

\begin{document}
\maketitle
\begin{abstract}
Several recent results show the influence of social contacts to spread
certain properties over the network, but others question the
methodology of these experiments by proposing that the measured
effects may be due to homophily or a shared environment.  In this
paper we justify the existence of the social influence by considering
the temporal behavior of Last.fm users.  In order to clearly
distinguish between friends sharing the same interest, especially
since Last.fm recommends friends based on similarity of taste, we
separated the timeless effect of similar taste from the temporal
impulses of immediately listening to the same artist after a friend.
We measured strong increase of listening to a completely new artist in a few hours period after a friend compared to non-friends representing a simple trend or external influence.
In our experiment to eliminate network independent elements of taste, we improved collaborative
filtering and trend based methods by blending with simple time aware recommendations based on the
influence of friends. Our experiments are carried over the two-year
``scrobble'' history of 70,000 Last.fm users.
\end{abstract}

\section{Introduction}

Several results show the influence of friends and contacts to spread obesity \cite{christakis2007spread}, loneliness \cite{cacioppo2009alone}, alcohol consumption \cite{rosenquist2010spread}, religious belief \cite{stroope2011social} and many similar properties in social networks.
Others question the methodology of these experiments \cite{lyons2011spread} by proposing that the measured effects may be due to homophily \cite{mcpherson2001birds}, the fact that people tend to associate with others like themselves, and a shared environment also called confounding or contextual influence.

Part of the appeal of Web 2.0 is to find other people who share similar interests.  Last.fm organizes its social network around music recommendation: users may automatically share their listening habits and at the same time grow their friendship.  Based on the profiles shared, users may see what artists friends really listen to the most.  Companies such as Last.fm use this data to organize and recommend music to people.

In this paper we exploit the timely information gathered by the Last.fm service on users with public profile to investigate how members of the social network may influence their friends' taste.  Last.fm's service is unique in that we may obtain a detailed timeline and catch immediate effects by comparing the history of friends in time and comparing to pairs of random users instead of friends.

Our contribution to the dispute on whether social contacts influence one another or whether the observed similarity in taste and behavior is only due to homophily, we show a carefully designed experiment to subtract external effects that may result in friends listening to similar music.
Homophily is handled by collaborative filtering, a method that is capable of learning patterns of similarity in taste without using friendship information.
Another possible source for users listening to the same music may come from traditional media: news, album releases, concerts and ads.  While the sources are hard to identify, common in them is that they cause temporal increase in popularity for the targeted artist.  These effects are filtered by another method that measures popularity at the given time and recommends based on the momentary popularity.

We blend collaborative filtering and temporal popularity recommenders with a method for influence prediction that we describe in this paper.  We consider events where a user listens to an artist for the first time closely after a friend listened to the same artist.
We obtain a 4\% of increase in recommendation quality, a strong result in view of the three-year Netflix Prize competition \cite{netflix-prize} to improve recommender quality by 10\%.  Note that we only give a single method that results in a stable strong improvement over the baselines. 

Our new  method is a lightweight recommender based on friends' past items that can be very efficiently computed even in real time.  Part of the efficiency comes from the fact that potential items from influencing friends are relative rare.  For this reason, the method in itself performs worse than the baselines, however it combines very well with them.  Indeed, influence based predictions improve the accuracy of a traditional factor model recommender by nearly as much as measuring popularity at the given time, a prediction that is strong in itself.  The fact that influences blend well prove that close events in the network bring in new information that can be exploited in a recommender system and also prove the existence of influence from friends beyond homophily.

\subsection{Related results}

The Netflix Prize competition \cite{netflix-prize} has recently
generated increased interest in recommender algorithms in the
research community and put recommender algorithms under a systematic thorough evaluation on standard data \cite{bell2007lnp}.  The final best results blended a very large number of methods whose reproduction is out of the scope of this paper.
As one of our baselines we selected a successful matrix factorization recommender described by Simon
Funk in \cite{funk06netflix} that is based on an approach reminiscent of gradient boosting \cite{friedman99gradient}. 

Closest to our results are the applications of network influence in collaborative filtering \cite{domingos2001mining}.  However in their data only ratings and no social contacts are given.  In another result \cite{goyal2010learning} over Flickr, both friendship and view information was present, but the main goal was to measure the strength of the influence and no measurements were designed to separate influence from other effects.

Bonchi \cite{bonchi2011influence} summarizes the data mining aspects of research on social influence. He concludes that ``another extremely important factor is
the temporal dimension: nevertheless the role of time in viral
marketing is still largely (and surprisingly) unexplored'', an aspect that is key in our result.

Since our goal is to recommend different artists at different times, our evaluation must be based on the quality of the top list produced by the recommender. This so-called top-$k$ recommender task is known to be hard \cite{deshpande2004item}.  For a recent result on evaluating top-$k$ recommenders is found in \cite{cremonesi2010performance}.

Music recommendation is considered in several results orthogonal to
our methods that will likely combine well.  Mood data set is created
in \cite{hu2007creating}.  Similarity search based on audio is given
in \cite{knees2007music}.  Tag based music recommenders
\cite[and many more]{eck2007automatic,tso2008tag}, a few of them based on Last.fm tags, 
use annotation and fall into the class of content based methods as
opposed to collaborative filtering considered in our paper.  Best
starting point for tag recommendation in general are the papers
\cite{jaschke2007tag,marlow2006ht06,markines2009evaluating}.  Note
that the Netflix Prize competition put a strong vote towards the
second class of methods \cite{pilaszy2009recommending}.

As a social media service, Twitter is widely investigated for
influence and spread of information.  Twitter influence as followers
has properties very different from usual social networks
\cite{kwak2010twitter}.  Deep analysis of influence in terms of
retweets and mentions is given in \cite{cha2010measuring}.  Notion of
influence similar to ours is derived in
\cite{cha2008characterizing,bakshy2011everyone} for Fickr and Twitter
cascades, respectively.  Note that by our measurement the Last.fm data
contains only a negligible amount of cascades as opposed to Twitter or
Flickr.

\section{The Last.fm data set}
\label{sect:dataset}

\begin{figure}
  \begin{center}
    \includegraphics[width=7.5cm]{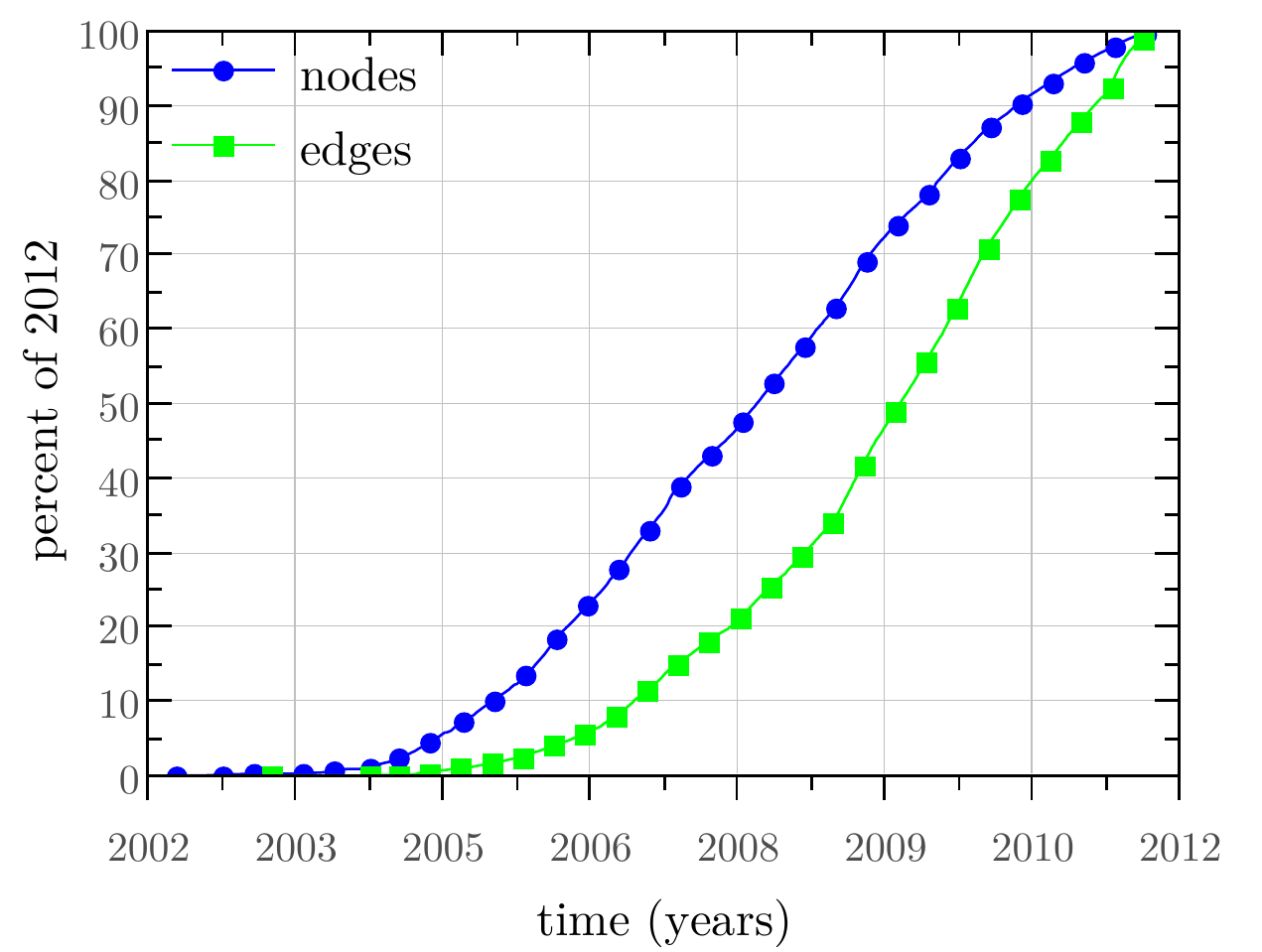}
    \caption{The number of the users and friendship edges in time as the fraction of the values at the time of the data set creation (2012).}
    \label{fig:networkGrowth}
  \end{center}
\end{figure}

\begin{figure}
  \begin{center}
    \includegraphics[width=7.5cm]{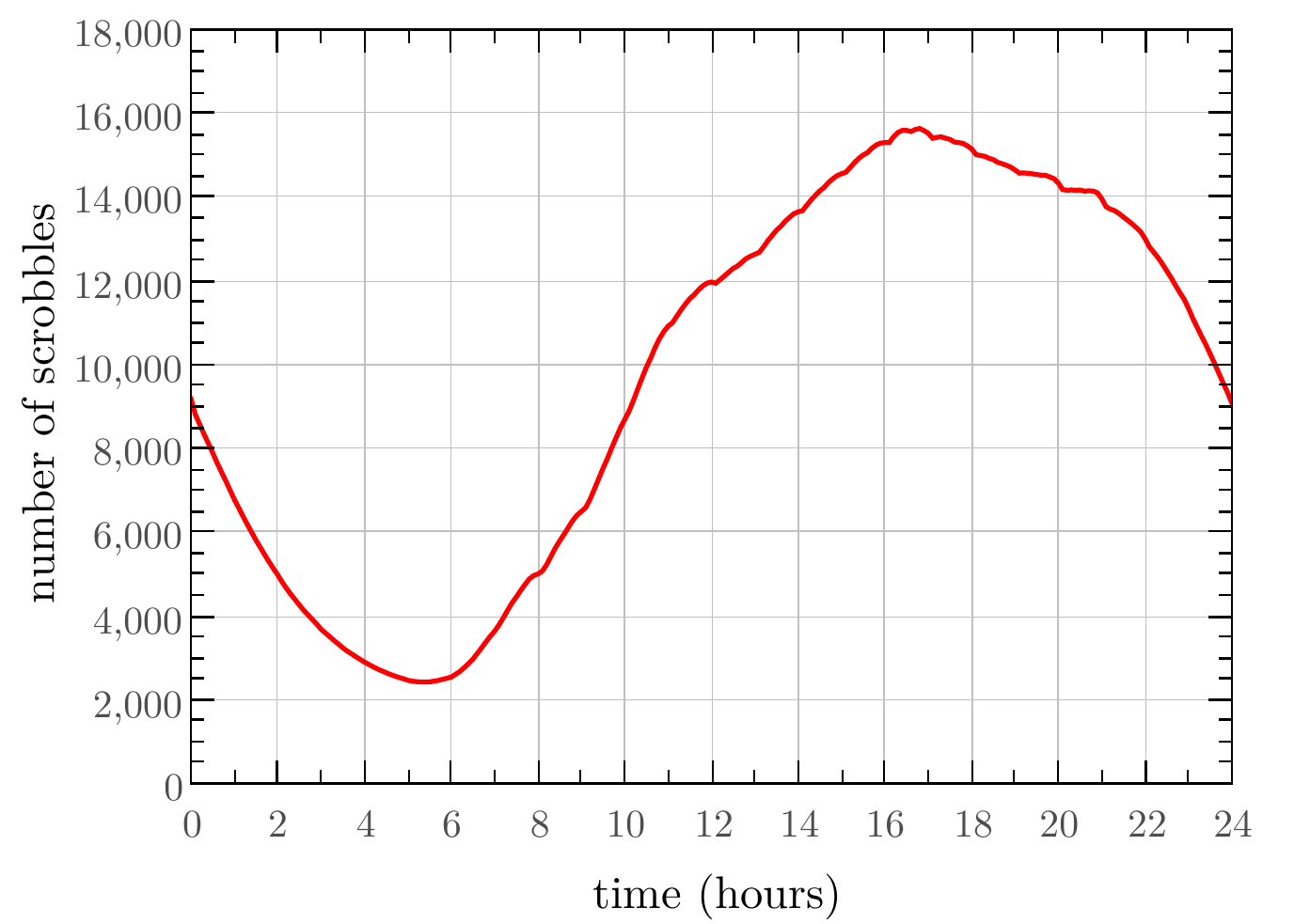} 
    \caption{Daily periodicity of scrobble count.}
    \label{fig:scrobble-periodicity}
  \end{center}
\end{figure}

Last.fm became a relevant online service in music based social networking. The idea of Last.fm is to create a recommendation system based on plugins nearly for all kind of music listening platforms. For registered users it collects, ``scrobbles''\footnote{The name ``scrobbling'' is a word by Last.fm, meaning the collection of information about user listening.} what they have listened. Each user has its own statistics on listened music that is shown in her profile. Most user profiles are public, and each user of Last.fm may have friends inside the Last.fm social network. Therefore one relevant information for the users is that they see their own and their friends' listening statistics.
We focus on two types of user information,
   \begin{itemize}
     \item the timeline information of users: user $u$ ``scrobbled''  artist $a$  at time $t$ ($u,a,t$),
     \item and the social network of users.
   \end{itemize}
Our data set hence consists of the contacts and the musical taste of the users.  Our goal is to justify the existence of the influence of social contacts, i.e.\ certain correlation the taste of friends in the user network. For privacy considerations, throughout our research, we selected an anonymous sample of users.  Anonymity is provided by selecting random users while maintaining a connected friendship network.  We set the following constraints for random selection:
\begin{itemize}
  \item User location is stated in UK;
  \item Age between 14 and 50, inclusive;
  \item Profile displays scrobbles publicly (privacy constraint);
  \item Daily average activity between 5 and 500.
  \item At least 10 friends that meet the first four conditions.
\end{itemize}
The above selection criteria were set to select a representative part of Last.fm users and as much as possible avoid users who artificially generate inflated scrobble figures. In this anonymized data set of two years of  artist scrobble timeline, edges of the social network are undirected and timestamped by creation date (Fig.~\ref{fig:networkGrowth}).  Note that no edges are ever deleted from the network.

\begin{figure}
  \begin{center}
    \includegraphics[width=7.5cm]{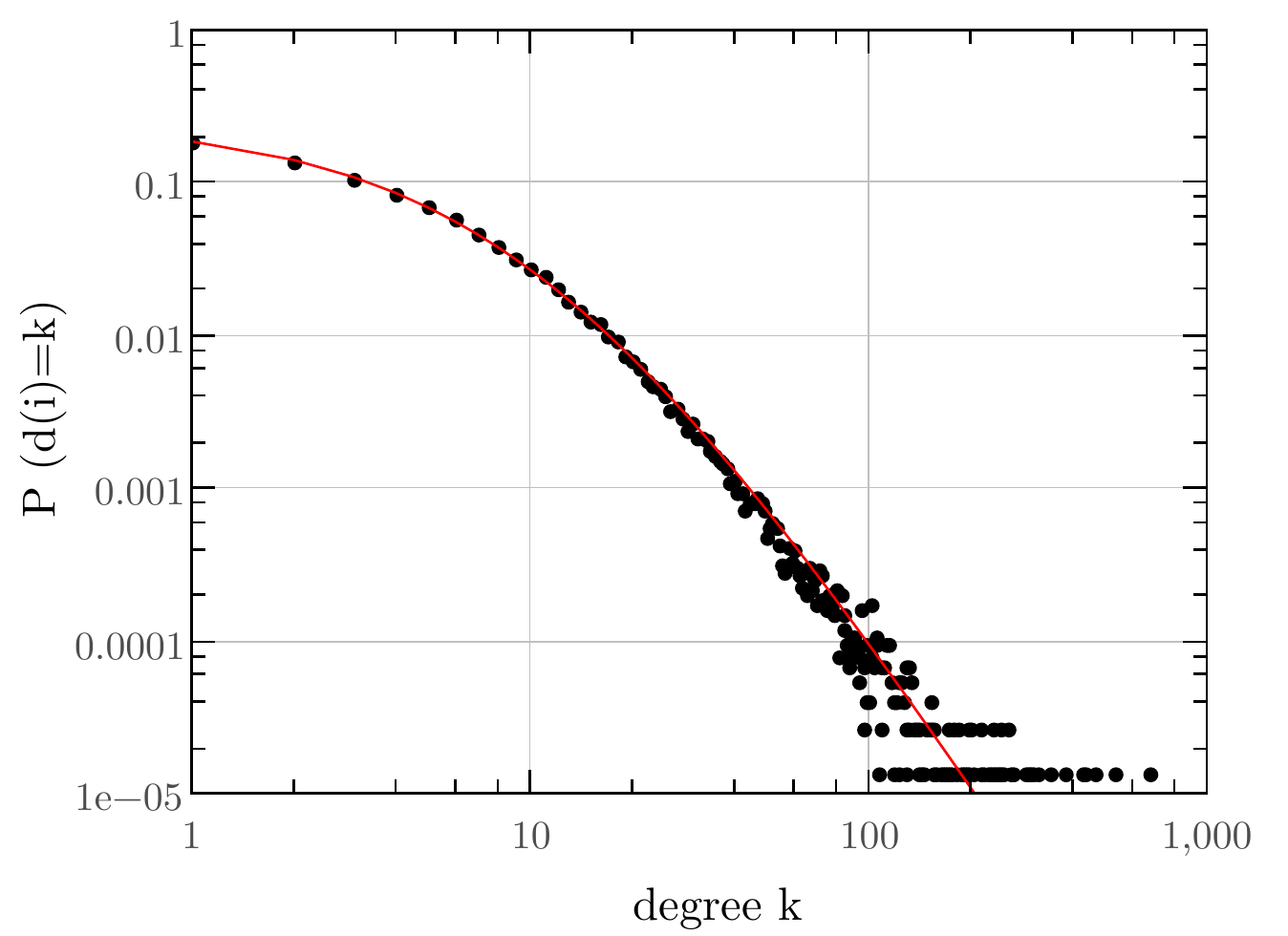}
    \caption{Degree distribution in the friendship network.}
    \label{fig:degDist}
  \end{center}
\end{figure}

The number of users both in the time series and in the network is 71,000 with 285,241 edges. The average degree is therefore 8, while the degree distribution follows shifted power-law as seen in Fig.~\ref{fig:degDist}
$$P(d(i)=k) \sim (x+s)^{-\alpha}$$
with exponent 3.8.

The time series contain 979,391,001 scrobbles from  2,073,395 artists and were collected between 01 January 2010 and 31 December 2011. Note that one user can scrobble an artist at different times. The number of unique user-artist scrobbles is 57,274,158. Fig.~\ref{fig:scrobble-periodicity} shows the daily fluctuations in the users scrobbling activity.

\section{Notion of network influence}
\label{sect:influence}

\begin{figure}
    \begin{center}
      \includegraphics[width=5 cm]{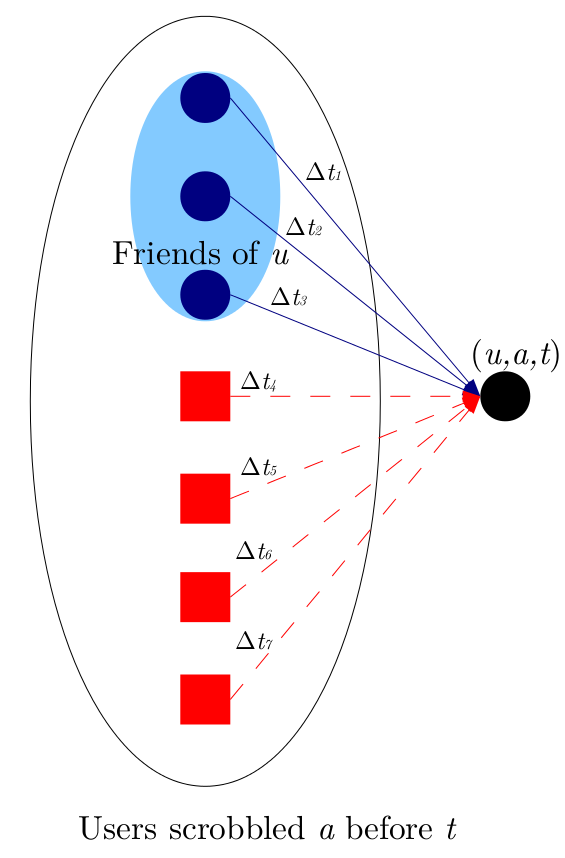}
      \caption{Potential influence on $u$ by some other user to scrobble $(u,a,t)$.}
      \label{fig:influence_figure}
    \end{center}
  \end{figure}

The key concept in this paper is a user $v$ \emph{influencing} another $u$ to scrobble $a$.  This happens if $u$ scrobbles artist $a$ the \emph{first time} at time $t$, after $v$ \emph{last scrobbling} the same artist at some time $t' < t$ before.  The time difference $\Delta t = t-t'$ is the \emph{delay} of the influence, as seen in Fig.~\ref{fig:influence_figure}.  Our key assumption is that, in the above definition, we observe influences between non-friends only by coincidence while some of the observed influence between friends is the result of certain interaction between them.  Our goal is to prove that friends indeed influence each other and this effect can be exploited for recommendations.

Similar influence definitions are given in \cite{goyal2010learning,cha2008characterizing,bakshy2011everyone}.
As detailed in \cite{bakshy2011everyone}, one main difference between these definitions is that in some papers $t'$ is defined as the first and not the last time when user $v$ scrobbles $a$. 

For smaller influence delay $\Delta t$, we are more certain that $u$ is affected by the previous scrobble of $v$.  The distribution of delay with respect to friends and non-friends will help us in determining the frequency and strength of influence over the Last.fm social network.
Each time user $u$ first scrobbles $a$, we compute the delay $\Delta t$ for all
users $v$ who scrobbled $a$ before $u$, if such users exist (see Fig.~\ref{fig:influence_figure}).

Out of the 57,274,158 first-time scrobbles of certain artist $a$ by some user, we find a friend who scrobbled $a$ before 10,993,042 times (19\%).
Note that one user can be influenced by more friends therefore the total number of influences is 24,204,977.
There is no influencing user for the very first scrobbler of $a$ in the data set.
For other scrobbles there is always an earlier scrobble by some other user, however that user may not be a friend of $u$.

\begin{figure}
    \centering
      \includegraphics[width=7.5cm]{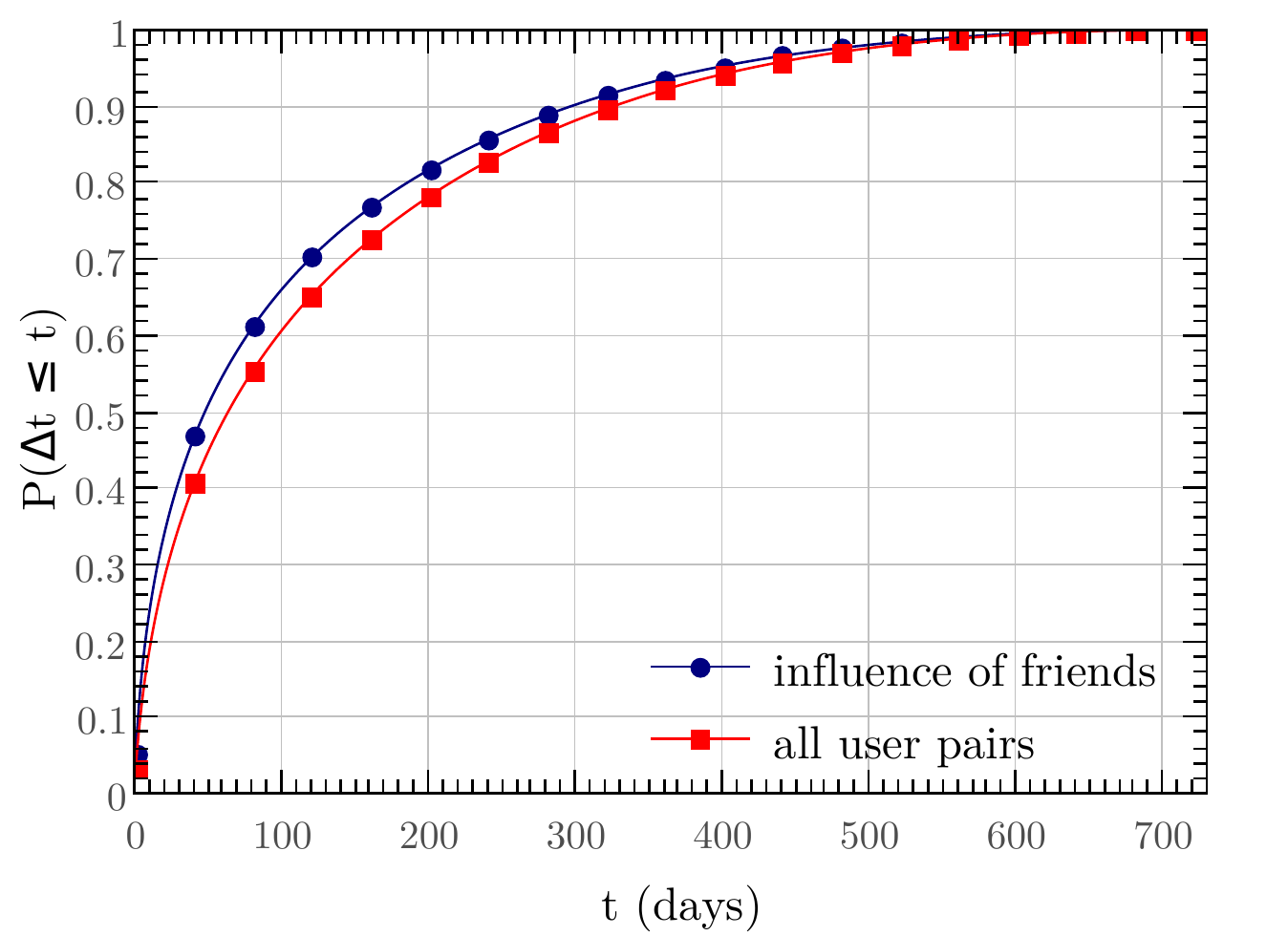}

      \includegraphics[width=7.5cm]{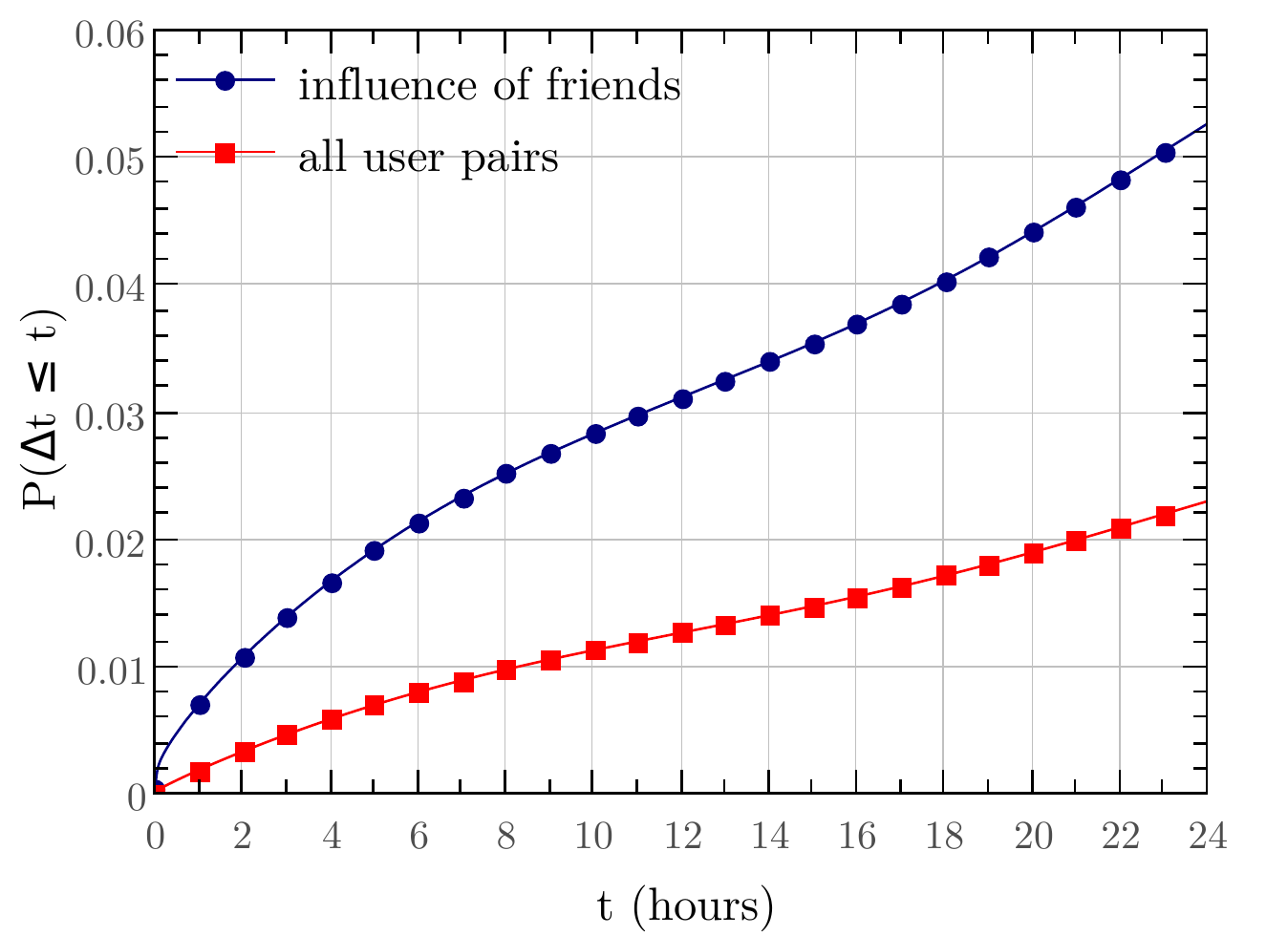}

    \caption{Fraction of influences with delay $\Delta t \le t$ as the function of $t$ as in \protect\eqref{eq:cdf}, in case of friends ($\CDF_F$) and non-friends ($\CDF_A$) over the entire timeline (\textbf{top}) and the first 24 hours (\textbf{bottom}).}
    \label{fig:tempinf}
  \end{figure}

Some of the observed influences may result by pure coincidence, especially when a new album is released or the popularity of the artist increases for some other reason.
In order to identify real influence, we compare the frequency of influence from friends and
from non-friends along delay $\Delta t$ as parameter.  We compute the
cumulative distribution function of all influences as a function of the delay,
\begin{equation}
\CDF_A (t) = \mbox{\vtop{\noindent fraction of influences with delay $\Delta t \le t$\\ among all influences.}}
\label{eq:cdf}
\end{equation}
Similarly, $\CDF_F(t)$ stands for the same function among influences between friends only.  Fig.~\ref{fig:tempinf} shows the functions for all users and friends.  The function of friends is above that of all users, i.e.\ we observe shorter
delay more frequently among friends.

\begin{figure}
  \begin{center}
    \includegraphics[width=7.5cm]{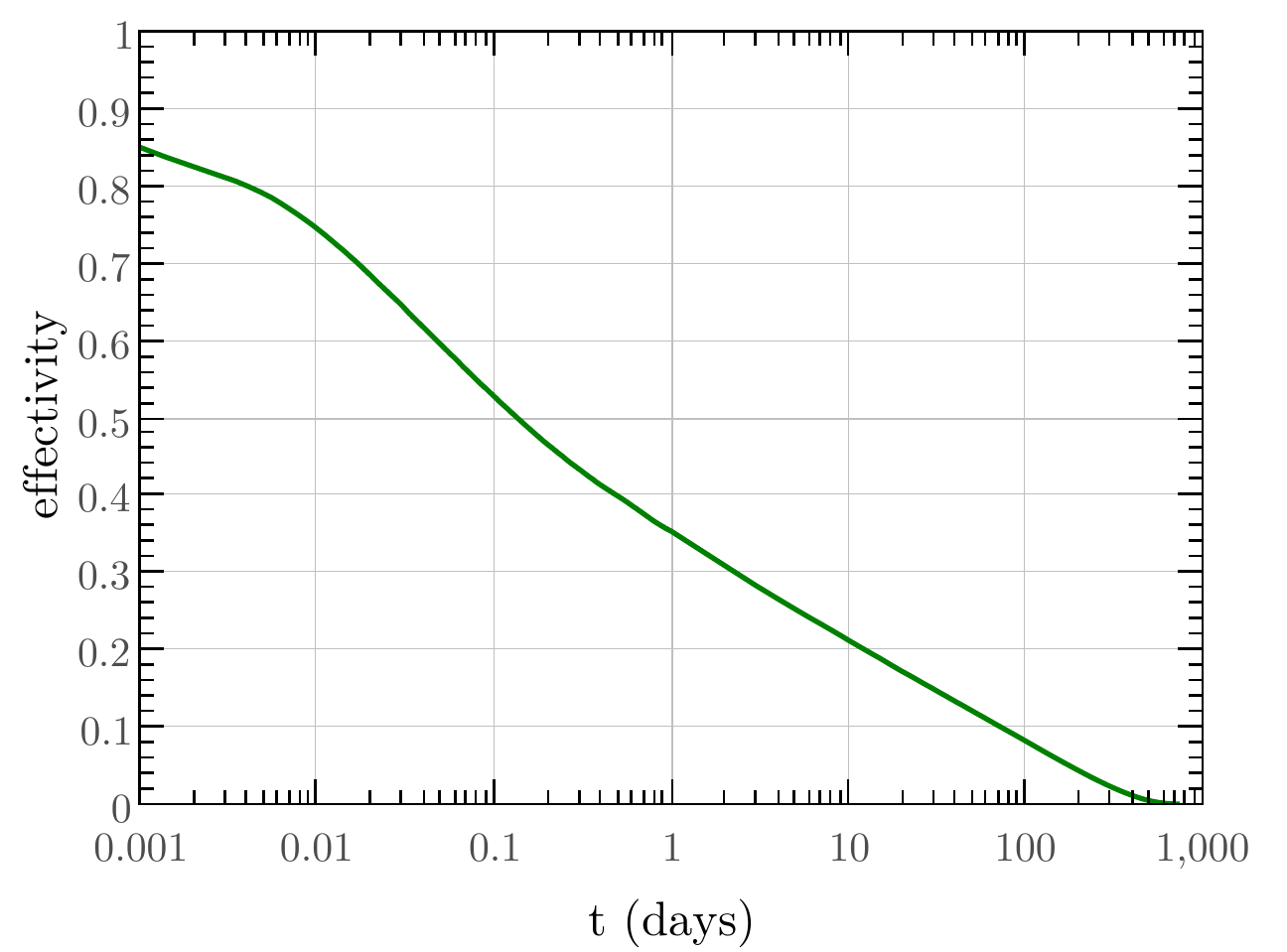}
    \caption{The measured effectivity of the influence (ratio of increase among friends compared to all users) as in \protect\eqref{eq:eff} very closely follows a logarithmic function of delay $\Delta t$.}
    \label{fig:effcurve}
  \end{center}
\end{figure}

Next we quantify the importance of friendship in influencing others as the \emph{effectivity} function.  The effectivity at $\Delta t$ is defined as the increase of influenced scrobbles among friends relative to all users that happen with delay at most $t$:
\begin{equation}
  \Eff(t)= \frac{\CDF_F(t) - \CDF_A(t)}{\CDF_F(t)}.
  \label{eq:eff}
\end{equation}
Fig.~\ref{fig:effcurve} shows the measured effectivity curve in the community. As expected, $\Eff(t)$ is a monotonically decreasing function of $t$. However, the decrease is slow unlike in some recent influence models that propose exponential decay in time~\cite{goyal2010learning}.  Therefore, we approximate $\Eff(t)$ with a slowly decreasing logarithmic function instead of an exponential decay.

\section{Influence based recommendation}
\label{sect:infRec}

\begin{figure}
  \begin{center}
    \includegraphics[width=9cm]{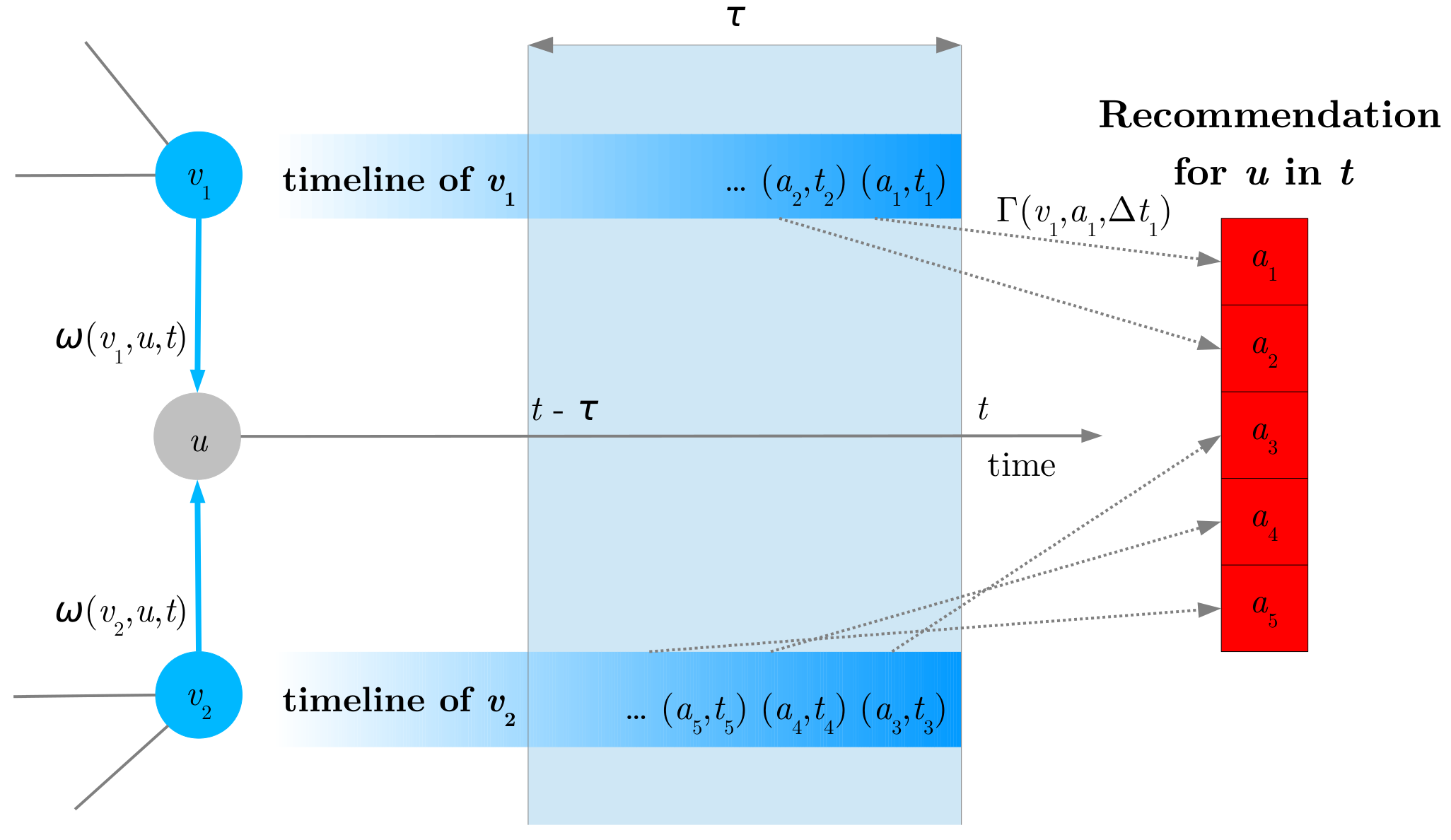}
    \caption{Scheme of the influence based recommender algorithm.}
    \label{fig:algorithm}
  \end{center}
\end{figure}

Next we use our notion of influence in the task of artist recommendation.  Influence depends on time and no matter how relative slow but the effectivity of a friend scrobbling an artist decays.  For this reason the influence based recommendation must be updated more frequently than traditional collaborative filtering methods. Also note that for a given user, our recommendation can be computed very efficiently by a pass over the recent history of friends.

Based on the measurements in the previous Section, we give a temporal network influence based recommender algorithm.  For a user $u$ at time $t$, we recommend based on friends' scrobbles before $t$.  The predicted score $\hat{r}(u,a,t)$ of an artist $a$ is based on a function $\Gamma$ of the time elapsed since the friend $v$ scrobbling $a$ (the delay $\Delta t$) and a function $\omega$ of the observed frequency of $v$ influencing $u$ in the past, as summarized in Fig.~\ref{fig:algorithm}.
Formally the predicted rating becomes
\begin{equation}
  \hat{r}(u,a,t)=\displaystyle\sum_{v \in n(u)} \Gamma (v,a,\Delta t)  \omega(v,u,t),
\end{equation}
where $n(u)$ denotes the friends of $u$, $\omega(v,u,t)$ is the strength of the influence between users $u$ and $v$, and $\Gamma(v,a,\Delta t)$ is the weight
between user $v$ and artist $a$ based on the delay.  

Our implementation depends on the two functions $\omega$ and $\Gamma$  defined in the next two subsections.  In an efficient algorithm, the value of $\omega$ can be stored in memory for all pairs of friends.  Alternately, $\omega$ can only be batch updated as the strength between two users are less time sensitive.  The values of $\Gamma$, however, depend on the actual time when the recommendation is requested.  As $\Gamma$ quickly decays with $\Delta t$, we only need to retrieve the past srobbles of all $v$, the friends of $u$.  This step can be efficiently implemented unless $u$ has too many friends.  In this latter case we could select only a few influencing friends based on the values of $\omega$, otherwise the recommendation is noisy anyway.  Our algorithm can hence be implemented even in real time.

\subsection{Influence as function of delay}
\label{sect:influencefunction}

The potential of influence decays as time elapses since the influencer $v$ scrobbled the given artist $a$. Based on the effectivity curve (see Fig.~\ref{fig:effcurve}) we approximate the strength of the influence with a monotonically decreasing logarithmic function
 \begin{equation}
   \Gamma(v,a,\Delta t)= 1 - C \cdot \log (\Delta t),
   \label{eq:predict}
 \end{equation}
where C is a global constant.

\subsection{Strength of influence between user pairs}

We recommend a recent scrobble by a friend by taking both the recency of the scrobble and the observed relation between the two users. For each pair of users $u$, the influenced and $v$, the influencer, we define the strength $\omega(v,u,t)$ as a step function in time as follows:
\begin{itemize}
  \item We initialize $\omega(v,u,0)=0$ for all pairs. 
  \item Assume that $u$ and $v$ become friends at time $t_0$.  We take a step and set $\omega(u,v,t_0)=\omega(v,u,t_0)=1$.
  \item If we observe an influence from $v$ to $u$ at time $t>t_0$ with time difference $\Delta t$, we take another step and increase $\omega(v,u,t)$ by
    \begin{equation}
    \omega(v,u,t) \leftarrow \omega(v,u,t)+\left ( 1 - C \cdot \log (\Delta t)\right ),
    \label{eq:learn}
    \end{equation}
    where $C$ is a global constant. For simplicity we use the same logarithmic function of the delay as in \eqref{eq:predict}. 
\end{itemize}

To speed up computations, we only consider influence with delay not more than a predefined time frame $\tau$. 
We apply $\tau$ for defining both $\omega$ in \eqref{eq:learn} and $\Gamma$ in \eqref{eq:predict} and hence in both cases we set
\begin{equation}
  C = 1/\log(\tau).
\end{equation}

\section{Real time recommendation evaluation}
\label{sect:measures}

Recommender systems in practice need to rank the best $k$ items for the user in real time.  In the so-called top-$k$ recommendation task \cite{deshpande2004item,cremonesi2010performance}, potentially we have to compute a new top list for every single scrobble in the test period.  The top-$k$ task is different from the standard recommender evaluation settings and needs carefully selected  metrics that we describe next.%in this Section~\ref{sect:measures}.  

Out of the two year scrobbling data, we use the full first year as training period.  The second year becomes the testing period where we consider scrobbles one by one.  We allow a recommender algorithm to use part or full of the data before the scrobble in question for training and require a ranked top list of artists as output.  We evaluate the given single actual scrobble $a$ in question against the recommended top list by computing the discounted cumulative gain with threshold $K$
  \begin{equation}
    \mbox{DCG@K}(a) = 
         \begin{cases}
               0 & \mbox{ if rank }(a) > K; \\ 
               \frac{\displaystyle 1}{\displaystyle \log_2 (\mbox{rank}(a)+1)} & \mbox{otherwise}.
         \end{cases}
    \label{eq:dcg}
  \end{equation}
Note that in this unusual setting there is a single relevant item and hence for example no normalization is needed as in case of the NDCG measure.  Also note that the DCG values will be small since the NDCG of a relative short sequence of actual scrobbles will roughly be equal to the sum of the individual DCG values.   The DCG measured over 100 subsequent scrobbles of different artists cannot be more than the ideal DCG, which is $\sum_{i=1}^{100} 1/\log_2 (i+1) = 20.64$ in this case (the ideal value is 6.58 for $K=20$). Hence the DCG of an individual scrobble will on average be less than 0.21 for $K=100$ and 0.33 for $K=20$. 

\begin{figure}
  \begin{center}
    \includegraphics[width=7.5cm]{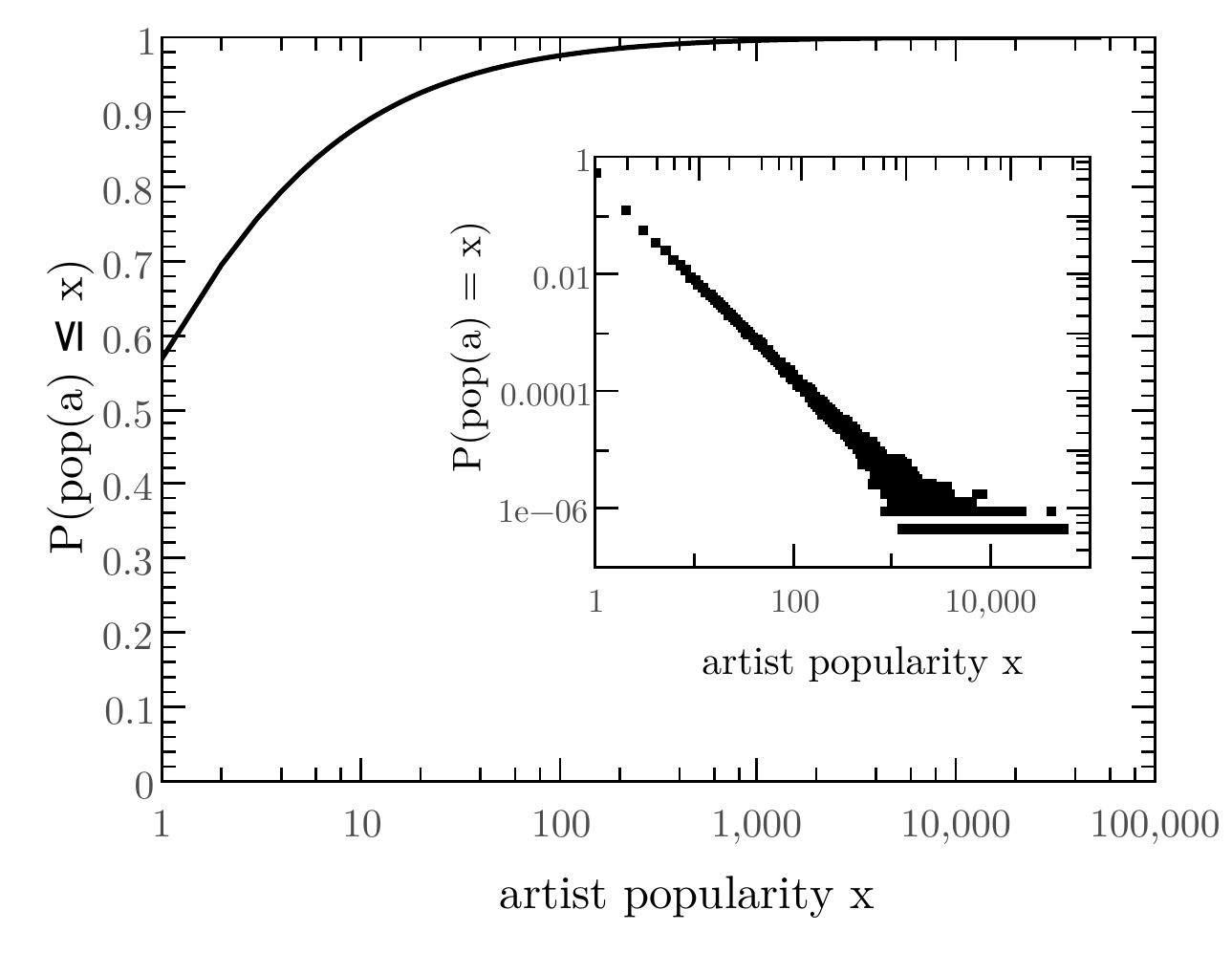}
    \caption{Distribution of scrobble count to a given artist and the cumulative distribution.}
    \label{fig:artistPop}
  \end{center}
\end{figure}

\begin{figure}
  \begin{center}
    \includegraphics[width=7.5cm]{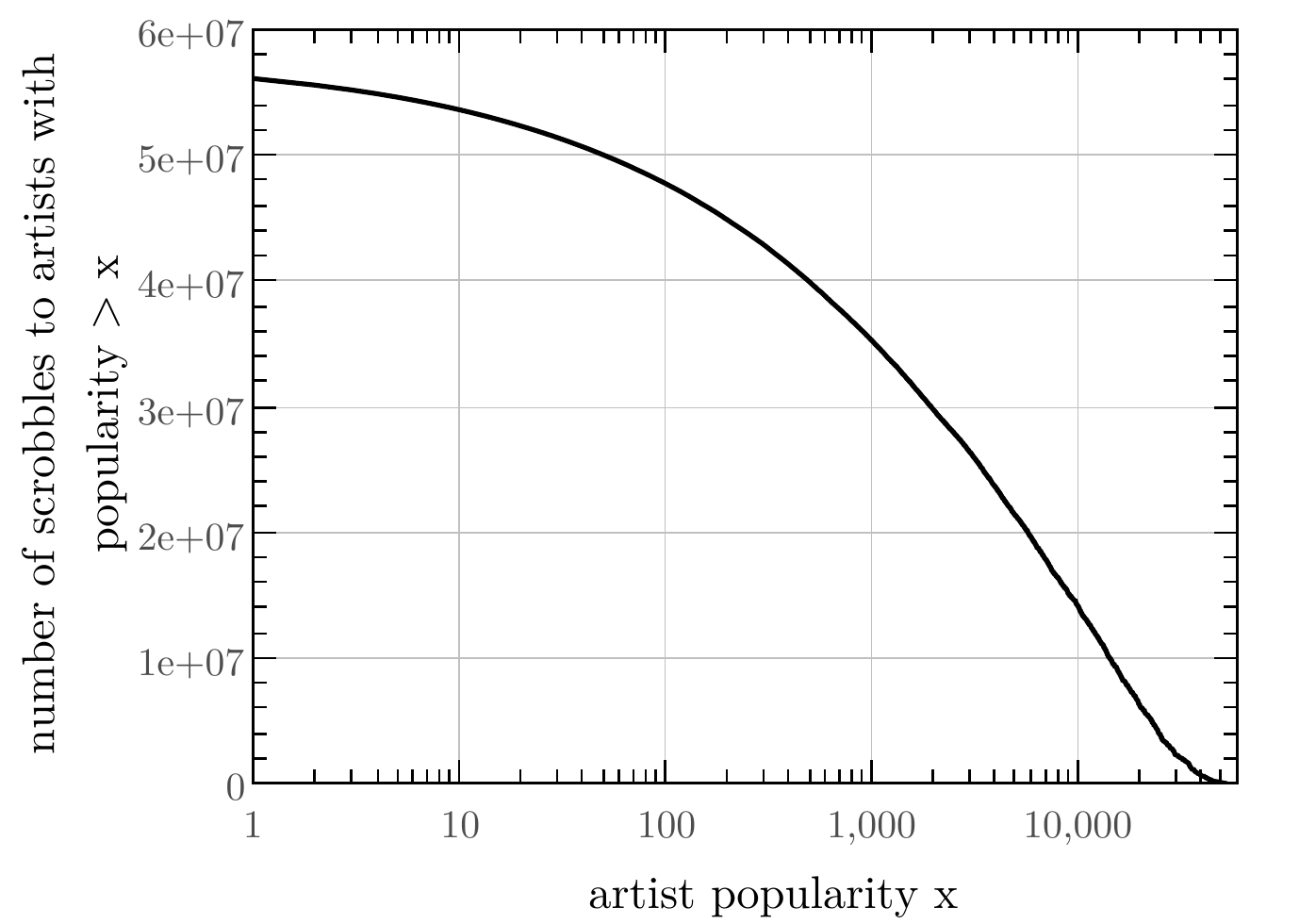}
    \caption{Fraction of scrobbles for artists with popularity at least a given value $x$, as the function of $x$.}
    \label{fig:artistPop4}
  \end{center}
\end{figure}

\begin{figure}
  \begin{center}
    \includegraphics[width=7.5cm]{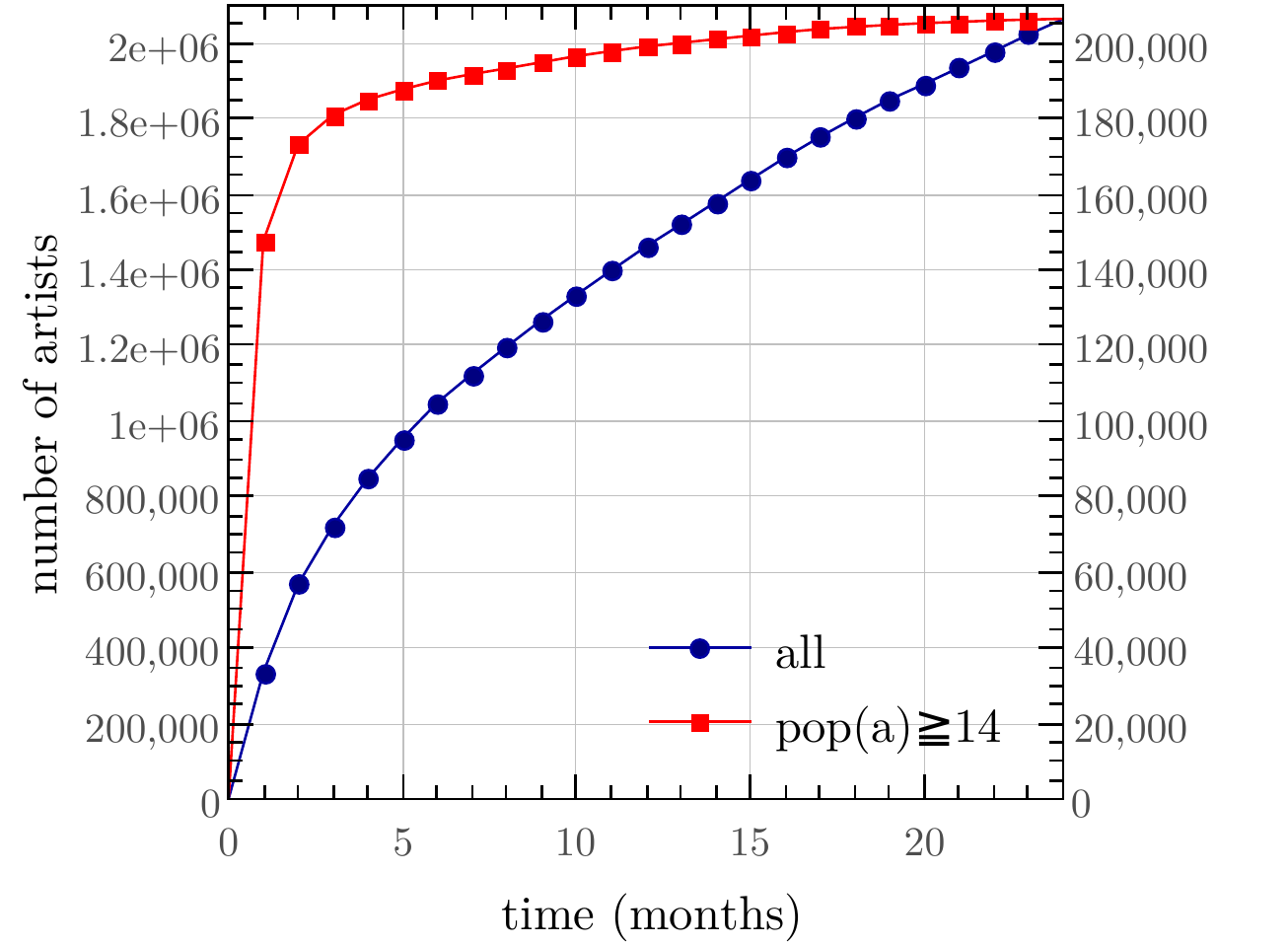}
    \caption{The number of different artists scrobbled before a given time in the two year period of the data set.}
    \label{fig:artistGrowth}
  \end{center}
\end{figure}

In our evaluation we discard infrequent artists from the data set both for efficiency considerations and due to the fact that our item based recommenders will have too little information on them.  As seen in Fig.~\ref{fig:artistPop}, the number of artists with a given scrobble count follow a power law distribution with near 60\% of the artists appearing only once.  While 90\% of the artists gathered less than 20 scrobbles in two years, as seen in Fig.~\ref{fig:artistPop4}, they attribute to only less than 10\% of the data set.  In other words by discarding a large number of artists, we only loose a small fraction of the scrobbles.  For efficiency we only consider artists of frequency more than 14. 

As time elapses, we observe near linear increase in the number of artists that appear in the data set in Fig.~\ref{fig:artistGrowth}.  This figure shows artists with at least 14 scrobbles separately.  Their count grows slower but still we observe a large number of new artist that appear in time and exceed the minimum count of 14.  Very fast growth for infrequent artists  may be a result of noise and unidentified artists from e.g.\ YouTube videos and similar Web sources.

\begin{figure}
  \begin{center}
    \includegraphics[width=7.5cm]{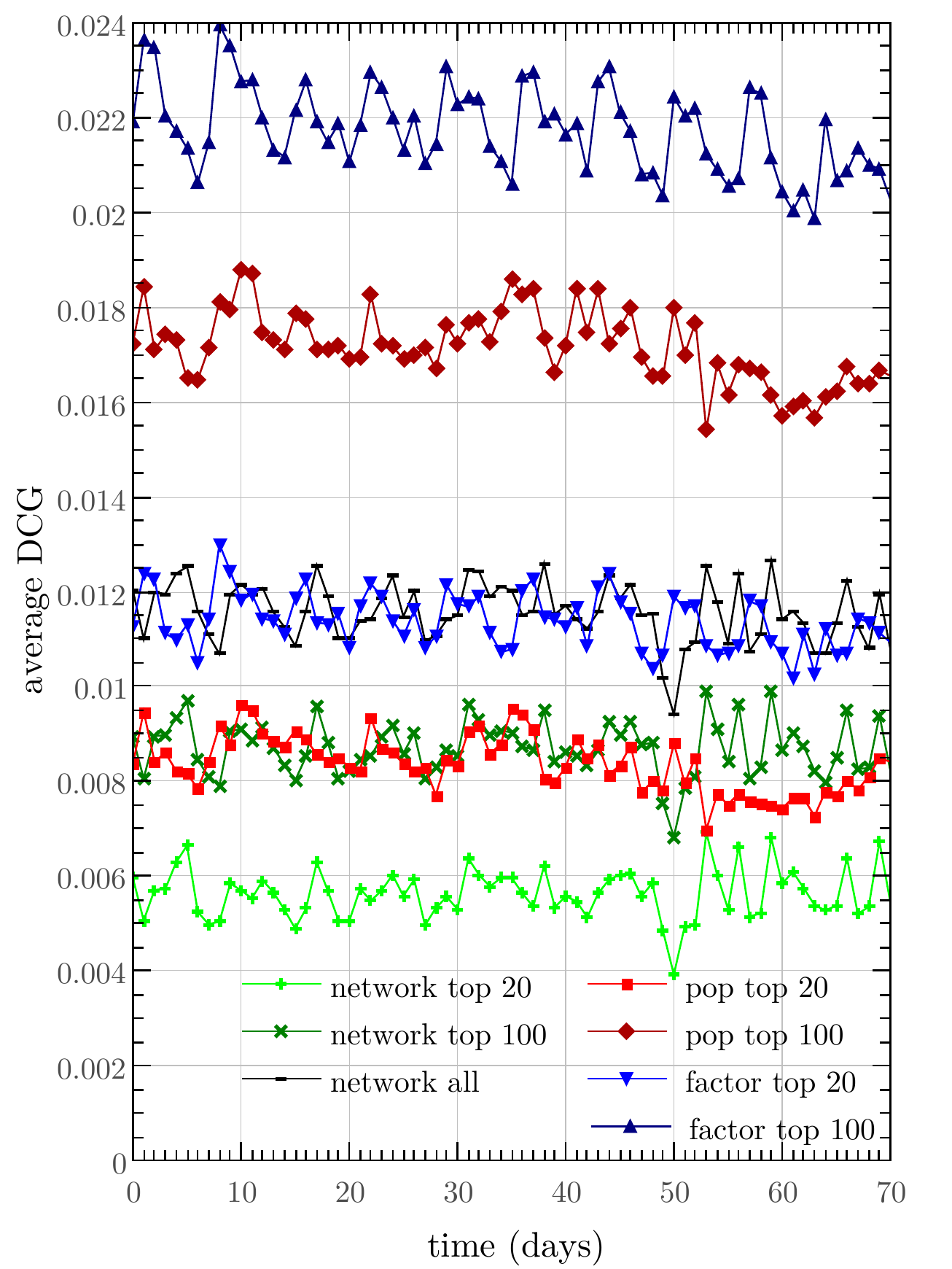}
    \caption{Daily average DCG@K as in \protect\eqref{eq:dcg} in a 70-day sample of the test period.  We show the three basic methods, from strongest to weakest, the factorization, temporal popularity, and the network influence recommenders.  For $K$ we measure two values, 20 and 100, except for network influence where we also show $K=\infty$ as the entire ranked list can be efficiently computed in this case.}
    \label{fig:dcg-time}
  \end{center}
\end{figure}

\begin{figure}
  \begin{center}
    \includegraphics[width=7.5cm]{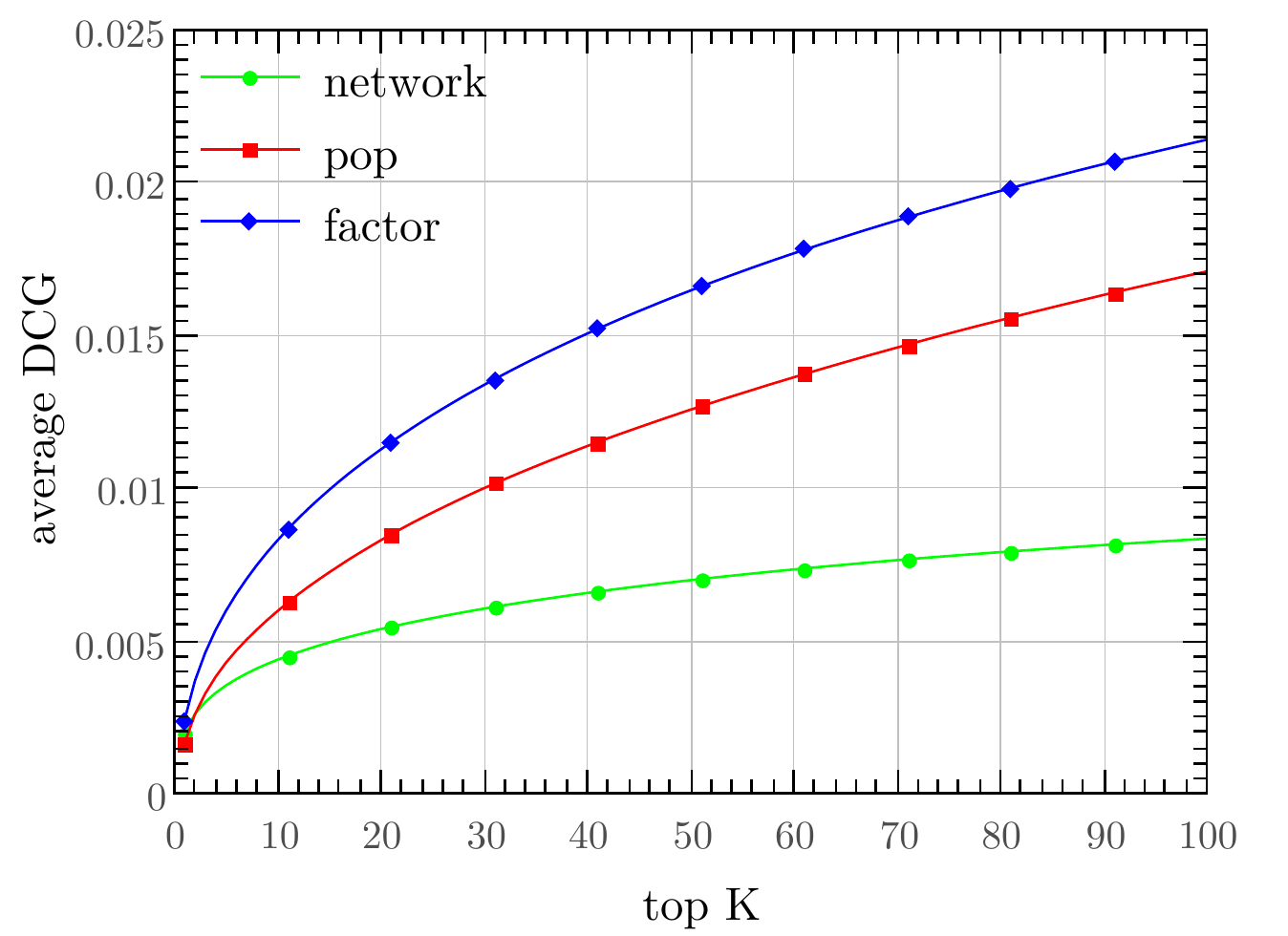}
    \caption{DCG@K as the function of $K$ for the three basic algorithms, for a time window $\tau$ equal to one week.}
    \label{fig:dcg-k}
  \end{center}
\end{figure}

\begin{figure}
  \begin{center}
    \includegraphics[width=7.5cm]{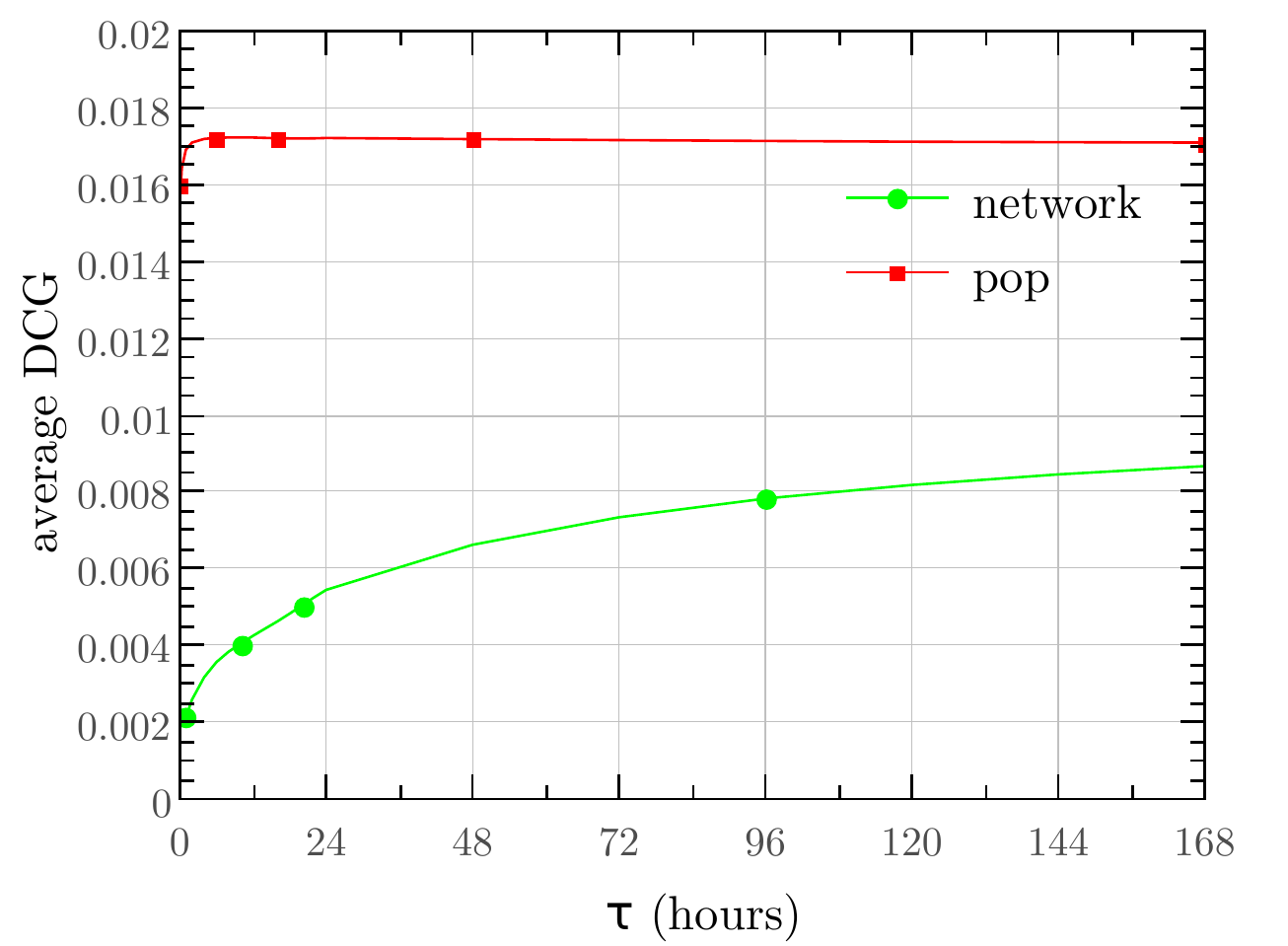}
    \caption{DCG@100 defined by equation \protect\eqref{eq:dcg} as the function of the time window threshold $\tau$ as in Section~\ref{sect:influencefunction}.}
    \label{fig:timeFrames}
  \end{center}
\end{figure}

\section{Music Recommendation Baseline Methods}

We describe one baseline method based on dynamic popularity in Section~\ref{sect:temporal} and one based on factorization in Section~\ref{sect:svd}. 

\subsection{Dynamic popularity based recommendation}
\label{sect:temporal}

Given a predefined time frame $\tau$ as in Section~\ref{sect:infRec}, at time $t$ we recommend an artist based on the popularity in time not earlier than $t-\tau$ but before $t$.  
In our algorithm we update the counts and store artists sorted by the current popularity.  In one time step we may either add a new scrobble event or remove the earliest one, corresponding to a count increment or decrement.  For globally popular items the sorted order can be maintained by a few changes in the order only.  To speed up the procedure, we may completely ignore part of the long tail and for others update the position only after a sufficiently large change in count. As future work we could also consider bursts and predict the popularity increase or decrease.

\subsection{Factor model based recommendation}
\label{sect:svd}

For our factor model based recommender we selected the implementation of Funk \cite{funk06netflix}.  In the testing period  we trained weekly models based on all data before the given week.  For each user, we constructed three times as many negative training instances as positive by selecting random artists with probability proportional to their popularity in the training period.
Each testing period lasted one week.  For each user, we compute a top list of predictions once for the entire week and evaluate against the sequence of scrobbles in that week. 

\section{Experiments}
\label{sect:eval}

First we give the daily average DCG@K defined by equation \protect\eqref{eq:dcg} in the second year testing period for the influence based and the two baseline recommenders.  Parameter $K$ in equation \eqref{eq:dcg} controls the length of the top list considered for evaluation.  In other words, $K$ can be interpreted as the size of the list presented to the user.  Practically $K$ must be small in order not to flood the user with information.  The performance of the three basic methods is shown in Fig.~\ref{fig:dcg-time} for $K=20$ and 100 and a time window $\tau$ in Section~\ref{sect:influencefunction} equal to one week.

The dependence on the top list size $K$ is measured in Fig.~\ref{fig:dcg-k} for $K\le100$.  We observe that our influence based method saturates the fastest.  This is due to the fact that the number of items recommended to a given user is usually small unless the user has a large number of very active friends.  For this reason we give blending results not just for the value $K=20$ that we consider practically feasible but also for 100 for comparison. 

Next we investigate the parameters of the individual algorithms. 
For a matrix factorization based method we use Funk's algorithm \cite{funk06netflix} with the following parameters that turned out to perform best in our experiments: learning rate${}=0.001$, feature number${}=20$, and initial feature value${}=0.1$.
We re-train the algorithm each week based on all past data.  For this reason we see weekly periodicity in the 10-week timeline of Fig.~\ref{fig:dcg-time}:  the factor model performs best immediately after the training period and slowly degrades in the testing period.

\begin{figure}
  \begin{center}
       \includegraphics[width=7.5cm]{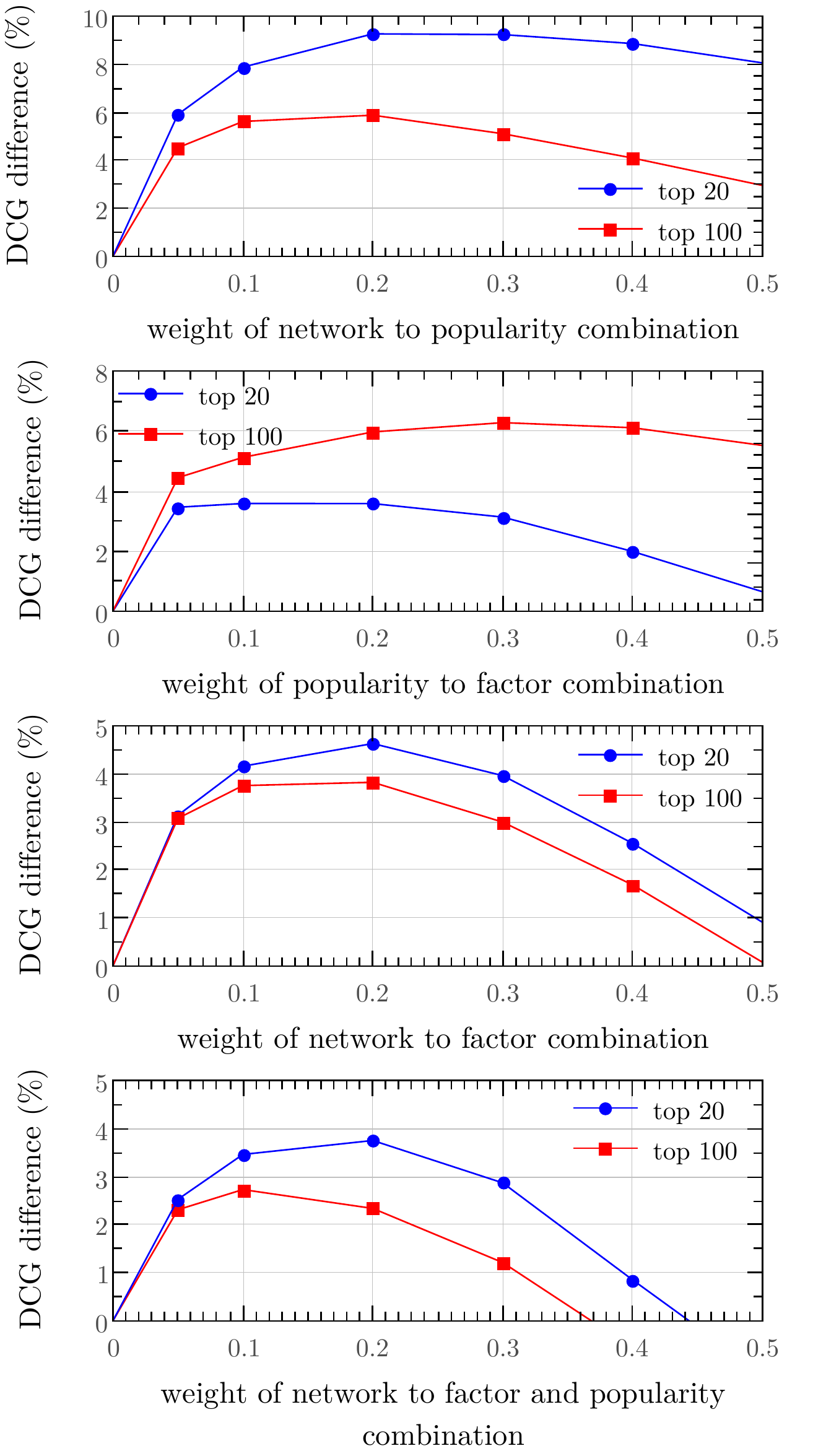}
    \caption{Blending DCG@K defined by \protect\eqref{eq:dcg} as the function of the linear combination weight.  From top to bottom: network influence and factor model; temporal popularity and factor model; network influence and factor model; finally network influence and the strogest combination of factor with popularity.}
    \label{fig:combine-fraction}
  \end{center}
\end{figure}

\begin{figure}
  \begin{center}
    \includegraphics[width=7.5cm]{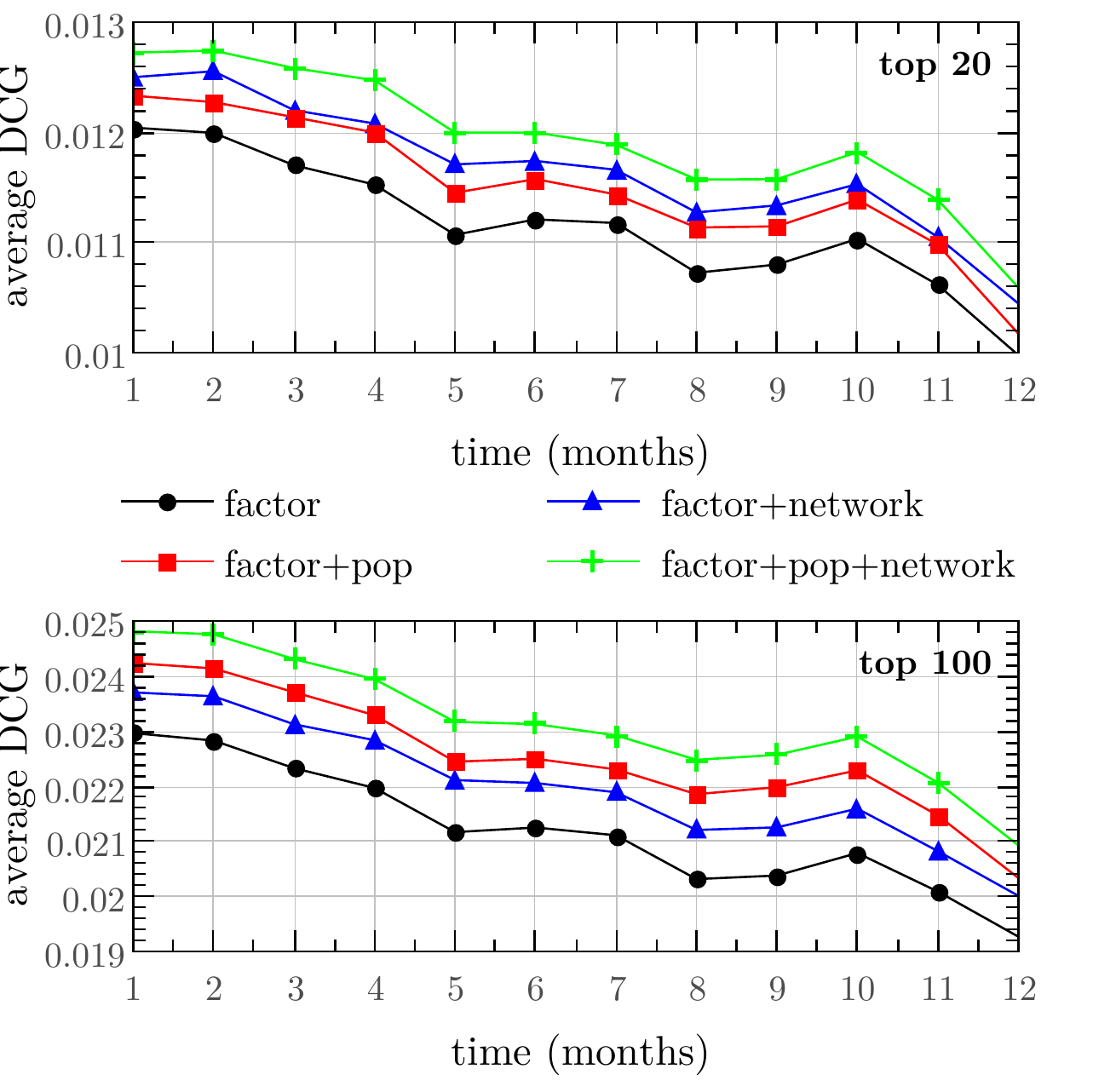}
    \caption{Monthly average DCG@20 (\textbf{top}) and DCG@100 (\textbf{bottom}) as defined by \protect\eqref{eq:dcg} in the test period for the factor model and its combinations.} 
    \label{fig:combine-timeline}
  \end{center}
\end{figure}

The popularity and influence based methods depend on the time frame:
the longer we look back in time, the more artists we can recommend.
If we carefully set the rank as a function of time, wider time frames
are advantageous for quality but put extra computational load.  For
the influence recommender $\tau$ is the maximum delay $\Delta t$ that
we consider as influence while for the popularity one $\tau$ is the
time interval that we use for frequency computation. We ran
measurements in the second year test period with different time frames
$\tau$ and computed the average DCG performance of the recommender
systems. Figure~\ref{fig:timeFrames} shows the average DCG scores with
different time frames.  The performance only slowly increases for time
frames longer than a day.  In what follows we set $\tau$ to be one week.

The final conclusion of the experiments is drawn by blending the three recommenders as shown in Figs.~\ref{fig:combine-fraction}--\ref{fig:combine-timeline}.  In our experiments we obtained the best results by linearly combining 1/rank instead of the predicted score. As an advantage of 1/rank, we need no score normalization.

Figure~\ref{fig:combine-fraction} shows the relative improvement of the recommenders as the function of the blending weights. After blending the recommenders pairwise, we selected the strongest popularity-factor combinations (3:7 and 2:8) and blended it with the network recommender. One can see that the influence recommender not only improves the results of the factor and popularity  recommenders, but combines well with their best blended result: the combination of the three methods outperforms the best blend of the factor and popularity models both for DCG@20 and DCG@100. The improvement is roughly 4\%. Figure~\ref{fig:combine-timeline} shows the monthly average DCG@20 and DCG@100 curves in the testing period in case of the different blended recommenders. Each curve shows the result of the best combination of the corresponding recommenders. In each case we observe stable improvement over the entire testing period.

\section*{Conclusions}

Based on a 70,000 sample of Last.fm users, we were able to measure the effect of certain user recommending an artist to her friends.  Our results confirm the existence of influence through the social network as opposed to the pure similarity of taste between friends.  We disproved the opinion that homophily could be the reason for friends listening to the same music or behave similarly by constructing a baseline that takes homophily and temporal effects into account. Over the baseline recommender, we achieved a 4\% improvement in recommendation accuracy when presenting artists from friends' past scrobbles that the given user had never seen before.  Our system has very strong time awareness: when we recommend, we look back in the near past and combine friends' scrobbles with the baseline methods.  The influence from a friend at a given time is certain function of the observed influence in the past and the time elapsed since the friend scrobbled the given artist.  In addition, our method can efficiently be computed even in real time.

For future work we plan to investigate whether the temporal social influence is specific to Last.fm dataset or can match to other kind of social network, e.g.\ Twitter.  We also plan to break down 
the analysis of influence spread by type of music, by age range, or by artist.

\section*{Acknowledgements}

To the Last.fm team for preparing us this volume of the anonymized data set  that cannot be efficiently fetched through the public Last.fm API.

\end{document}